\documentclass[fleqn,usenatbib]{mnras}

\usepackage[T1]{fontenc}

\DeclareRobustCommand{\VAN}[3]{#2}
\let\VANthebibliography\thebibliography
\def\thebibliography{\DeclareRobustCommand{\VAN}[3]{##3}\VANthebibliography}

\usepackage{graphicx}	
\usepackage{amsmath}	
\usepackage{amssymb}	

\newcommand{\be}{\begin{equation}}
\newcommand{\ee}{\end{equation}}
\newcommand{\beal}{\begin{aligned}}
\newcommand{\eeal}{\end{aligned}}
\usepackage{xcolor}

\title[Post-decoupling EM variability from CBDs]
{
Electromagnetic variability from circumbinary discs around binary black holes during their post-decoupling epoch
}

\author[R. Mignon-Risse et al.]{
Rapha\"{e}l Mignon-Risse,$^{1,2,3,4}$\thanks{E-mail: raphael.mignon-risse@ntnu.no}\thanks{E-mail: raphael.mignon-risse@lam.fr}
Peggy Varniere,$^{5,6}$
Fabien Casse$^{5}$
\\
$^{1}$Department of Physics, Norwegian University of Science and Technology, NO-7491 Trondheim, Norway\\
$^{2}$Universit\'e Paris Cit\'e, CNRS, CNES, Astroparticule et Cosmologie, F-75013 Paris, France\\
$^{3}$Department of Theoretical Physics, Atomic and Optics, Campus Miguel Delibes, University of Valladolid, Paseo Bel\'en, 7, \\ 47011, Valladolid, Spain \\
$^{4}$Aix-Marseille Universit\'e, CNRS, CNES, LAM, Marseille, France\\
$^{5}$Universit\'e Paris Cit\'e, CNRS, Astroparticule et Cosmologie, F-75013 Paris, France\\
$^{6}$Universit\'e Paris-Saclay, Universit\'e Paris Cit\'e, CEA, CNRS, AIM, 91191, Gif-sur-Yvette, France
}

\date{Accepted XXX. Received YYY; in original form ZZZ}

\pubyear{2025}

\begin{document}
\label{firstpage}
\pagerange{\pageref{firstpage}--\pageref{lastpage}}
\maketitle

\begin{abstract}
We present general-relativistic hydrodynamical simulations of {inviscid circumbinary discs (CBDs) around} {near equal-mass} binary black holes (BBH) in the binary-disc post-decoupling epoch.
We use an approximate BBH spacetime {with a} post-Newtonian inspiral {motion trajectory} from ${\sim}80 (M/10^7 \mathrm{M_\odot}) \, \mbox{days}$ (separation of ${\sim}\,30$ gravitational radii) to ${\sim}100 (M/10^7 \mathrm{M_\odot}) \, \mbox{minutes}$ before merger.
{Initial data for the inspiral runs are produced from circular-orbits runs} covering the formation {timescale} of the overdens{e} {\lq}lump{\rq}, orbiting the CBD inner edge.
The CBD non-axisymmetries (spiral waves and lump) lead to non-negligible angular momentum transport with effective viscosity ${\alpha_\mathrm{eff} \, {\sim} \, 10^{{-3}}{- 2\times 10}^{-2}}$. 
We post-process these simulations with a general-relativistic ray-tracing code to obtain synthetic observations in thermal emission.
We find the lump and its associated {electromagnetic (EM)} modulation, already reported in the pre-decoupling epoch, to survive post-decoupling up until the end of the simulation. 
For LISA sources, our findings point to an active EM signature in UV during optimal gravitational wave source localization.
For PTA sources and current BBH candidates detected through their optical periodicity: the lump{ in a low-viscosity CBD is a possible, though not unique, origin} for the observed periodicity. 
\end{abstract}




\section{Introduction}

Supermassive black holes are believed to reside in most galactic nuclei \citep{ferrarese_supermassive_2005}, so galaxy merger should trigger, at some point, the formation of supermassive binary black holes (BBHs) in gas-rich environments.
A combination of mechanisms is thought to extract angular momentum from the BBH to decrease its orbital separation, through gravitational interactions with the surrounding stars and gas, until gravitational wave (GW) emission becomes the dominant process \citep{milosavljevic_final_2003}.
At this stage, the flow morphology around the pre-merger BBH is expected to take the form of a circumbinary disc (CBD, \citealt{artymowicz_mass_1996}) feeding two individual discs (e.g. \citealt{gunther_circumbinary_2002})
As the inspiral motion accelerates, it eventually outpaces the inflow velocity from the CBD: this is the so-called binary-disc decoupling (\citealt{armitage_accretion_2002,liu_doubledouble_2003}) we refer to throughout this manuscript.
While many -- still on-going -- efforts have been put on predicting the pre-decoupling EM signatures of BBHs and using those in the current search for EM candidates \citep{dorazio_observational_2023}, the survival of these signatures in the post-decoupling phase is not clear yet.

Some BBH candidates have already been proposed based on their optical modulations detected over $3-5$ cycles (e.g. \citealt{graham_systematic_2015}). 
In the binary scenario, this optical modulation could be produced by the accretion rate variability (e.g. \citealt{farris_characteristic_2015}), the Doppler boost onto the orbiting mini-discs (e.g. \citealt{dorazio_relativistic_2015}) or by one of the CBD non-axisymmetries in the pre-decoupling phase, namely the orbiting lump and the two spiral arms {(\citealt{tang_late_2018}, }\citealt{mignon-risse_simple_2025}), each origin being associated with a different period.
If the observed modulation comes from the lump, several sources {should} have a separation ${r_{12}} \,{\lesssim}\,30$~M, e.g. SDSS J014350.13+141453.0 (Table 2 of \citealt{graham_systematic_2015}).
Indeed, for an observed lightcurve period $P_\mathrm{LC}$, binary mass $M$ and source redshift $z$, we can compute the separation (in units of $M$) as
\be
{r_{12,\mathrm{lump}}}\,  {\sim} \, 29 \left( \frac{P_\mathrm{LC}}{10^3 \mathrm{days}}  \right)^{2/3}  \left(  \frac{M}{10^9 \mathrm{M_\odot}}  \right)^{-2/3} \left( \frac{1+2}{1+z} \right)^{2/3} M.
\label{eq:blump}
\ee
{For this, we assumed} the lump is causing the modulation, the lump and orbital periods are linked by $\mathrm{P_{lump}} \, {\sim} \, 6 \, \mathrm{P_\mathrm{orb}}$ (e.g. \citealt{farris_characteristic_2015}, \citealt{mignon-risse_origin_2023}), and {that the BHs are in circular,} Keplerian orbit. 
At tens $M$ of separation, those could have already decoupled from their CBD depending on the disc's properties (e.g. \citealt{dittmann_decoupling_2023}).
This raises the question of the survival of pre-decoupling EM signatures, in particular that of the lump, in the post-decoupling phase: we explicitly focus on this phase here.
The same question is of interest for $10^{{4}-7}\, \mathrm{M_\odot}$, LISA-type BBHs, because localization will be optimal close to merger \citep{mangiagli_observing_2020}.

Most numerical efforts have focused on potential EM signatures linked to accretion rate variability (e.g. \citealt{farris_characteristic_2015}) or mini-discs, in the equal-mass case.
The accretion rate, as well as the mini-disc surface have been found to drop in the post-decoupling phase as the BBH inspiral accelerates (e.g. \citealt{dittmann_decoupling_2023}), suggesting a drop in the associated luminosity{, $1 (M/10^6 \mathrm{M}_\odot)$~day before merger (\citealt{krauth_disappearing_2023}, \citealt{franchini_emission_2024})}.
{In the post-decoupling epoch, the CBD is expected to be disconnected from the BBH so one may wonder if any EM variability such as the one directly induced by the lump may survive during this stage of the system (\citealt{bode_mergers_2012}, \citealt{gold_accretion_2014-1}, \citealt{farris_binary_2015}, \citealt{dittmann_decoupling_2023}, \citealt{krauth_disappearing_2023}, \citealt{franchini_emission_2024}).
The CBD emission variability would add up to aforementioned contributions, but at lower frequencies than the mini-discs, in the optical/UV band (e.g. \citealt{dascoli_electromagnetic_2018}, \citealt{tiwari_radiation_2025}).}

{Several studies have included an $\alpha-$disk model or the magneto-rotational instability (MRI, \citealt{balbus_powerful_1991}) to trigger angular momentum transport and thereby accretion (see Table~\ref{table:codes}).
Instead, here the angular momentum transport within the CBD is driven by the spiral waves, the lump and more generally, non-axisymmetries, without additional process (similar to \citealt{farris_binary_2011}); it can be somehow considered as the low-viscosity limit. 
Such a low-viscosity disc could form, for instance, following a tidal disruption event \citep{coughlin_tidal_2017}.
We choose BBH parameters prone to the lump formation in pre-decoupling epoch.
T}he mass ratio ($q\, {\equiv} \, M_1/M_2 \, {\le} \, 1$) is unlikely to be $1$ in the general case, and the lump formation becomes inhibited as $q$ decreases (\citealt{dorazio_accretion_2013}, \citealt{mignon-risse_origin_2023}): we will thus explore $q\, {=} \, 1$ and $q\, {=} \,0.3$.
{Moreover, the lump is reported in low-eccentricity ($e \, {\lesssim} \, 0.1$) systems (e.g. \citealt{siwek_preferential_2022}, \citealt{dorazio_fast_2024}): we will thus focus on non-eccentric orbits. Such circularization could be favored, close to merger, by GW emission.}
2D hydrodynamical (e.g. \citealt{farris_binary_2014}) and 3D magneto-hydrodynamical \citep{shi_three-dimensional_2012}, non-excised and excised simulations agree qualitatively and quantitatively on the flow morphology \citep{duffell_santa_2024}, even with different levels of complexity to deal with thermodynamical processes (e.g. isothermal in \citealt{shi_three-dimensional_2012} or with entropy-based cooling in \citealt{noble_circumbinary_2012}).
Our setup for this problem thus consists of 2D, inviscid GRHD simulations with an excised cavity, first in circular orbit (control) runs to set initial data for the inspiral runs and to compare with. 
{These are performed }in an approximate BBH metric (\citealt{ireland_inspiralling_2016}, see \citealt{mignon-risse_impact_2023}).
{While the incorporation of a GR metric, with an excised cavity, is not necessary to get most CBD features \citep{duffell_santa_2024}, this gives us the relativistic framework (e.g. Lorentz factors) to post-process these simulations } with a GR ray-tracing code in the same BBH metric {and} produce synthetic observations in thermal emission.
{Indeed, GR effects can have a non-negligible impact on the emerging radiation in single (e.g. \citealt{vincent_flux_2013}) and binary (e.g. \citealt{dorazio_relativistic_2015}, \citealt{dascoli_electromagnetic_2018}) BHs.}

{We adopt the geometric units $\mathrm{G} \,{=}\, \mathrm{c} \,{=}\, 1$, so the length unit is $r_\mathrm{g} \, {\equiv} \,\mathrm{G}M/ \mathrm{c^2}  \, {=} \, M$ and the time unit is $r_\mathrm{g} / \mathrm{c} \, {=} \, M$.}
Since the BBH mass scales out {in the post-Newtonian inspiral formulas}, we give timescale estimates based on $M \, {=} \, 10^7 \mathrm{M_\odot}$ which is at the transition between LISA and current optical BBH candidates or Pulsar Timing Arrays (PTAs) sources. 

\begin{center}
\begin{table}
\caption{Overview of the comparable studies on CBDs around BBHs incorporating GW-driven inspiral.
The column $\alpha$ gives the value of the $\alpha$ viscosity or the angular momentum transport model.
We deduced $\alpha$ from the kinematic viscosity for \citealt{dittmann_decoupling_2023}.
 \citealt{franchini_emission_2024} varied the Mach number and thus explored two regimes of kinematic viscosity at fixed $\alpha$ value.
$^\star$We computed an effective ${\alpha}$ viscosity in our setup (see the main text) {and indicate below the interval of values obtained}.
The column "Grav." refers to the treatment of the gravitational influence of the BBH onto the gas.}
\label{table:codes}  
\begin{tabular}{ | c | c | c | c | c } 
 \hline
 Reference & $q$ & $\alpha$ & 2D/3D & Grav.  \\  \hline
 \cite{noble_circumbinary_2012} & $1$ & MRI & 3D & GR \\ 
  \cite{farris_binary_2015} & $1$ & $0.1$ & 2D & Newt. \\ 
  \cite{tang_late_2018} & $1$ & $0.1$ & 2D & PN \\ 
  \cite{dittmann_decoupling_2023} & $1$ & $0.01-1$ & 2D & Newt. \\  
  \cite{krauth_disappearing_2023} & $1$ & $0.1-0.3$ & 2D & Newt. \\ 
   \cite{franchini_emission_2024} & $1$ & $0.1$ & 3D & Newt. \\ 
   This work & $0.3{-}1$ & $0.001 \, {-} \,0.0{2}^\star$ & 2D & GR \\ 
 \hline
\end{tabular}
\end{table}
\end{center}

\section{Numerical methods}
\label{sec:method}

Our goal is to question the survival of the {CBD} EM variability reported pre-decoupling, dominated by the orbiting lump, in the post-decoupling epoch for $q \in \{ 0.3,1 \}$.
We present here how our setup answers the numerical challenges to make this problem computationally feasible.

We use {\tt e-NOVAs} (\citealt{varniere_novas_2018}, \citealt{mignon-risse_impact_2023}) which computes the fluid evolution with {\tt GR-AMRVAC} \citep{casse_impact_2017} and solves the ray-tracing problem with {\tt GYOTO} \citep{vincent_gyoto_2011} to produce synthetic observations in thermal emission, in general relativity.

\subsection{General-relativistic hydrodynamics}

{Our simulations are two-dimensional spherical general-relativistic hydrodynamical, integrated over the direction perpendicular to the disc (orbital) plane and centered onto the center of mass of the BBH.
We solve for the local conservation of baryon number density, momentum and entropy over a background, approximate, analytical BBH metric valid in the circumbinary region, namely
\be
\beal
\partial_t ({\cal D}) + \partial_j \left[  {\cal D} \left( \alpha {\rm v}^j - \beta^j \right) \right] &= 0 ,\\
\partial_t ({\cal S}_i) + \partial_j \left(  \left[ {\cal S}_i (\alpha {\rm v}^j - \beta^j) + \alpha{\cal P}\delta_i^j \right] \right) &={\cal S}_j \partial_i \beta^j +  \\
  \frac{\alpha}{2} \left( {\cal S}^j {\rm v}^k + {\cal P}\gamma^{jk}\right)\partial_i \gamma_{jk} -&\sqrt{\gamma}(\Gamma^2\Sigma h - P)\partial_i \alpha  , \\
\partial_t ({\cal D S} ) + \partial_j \left[  {\cal D S } \left( \alpha {\rm v}^j - \beta^j \right) \right] &= 0,\\
\label{eq:eqs}
\eeal
\ee
where ${\cal D} \equiv \sqrt{\gamma}\  \Gamma\Sigma$ is the relativistic density, $\sqrt{\gamma}$ the square root of the spatial metric determinant, $\Gamma$ the Lorentz factor and $\Sigma$ the gas rest surface density;
${\cal S}_i \equiv \sqrt{\gamma}\ \Gamma^2\Sigma h{\rm v}_i$ is the relativistic momentum, with $h=1+\varepsilon + P/\Sigma$  the specific enthalpy, $v_i$ the covariant velocity, $\varepsilon$ the gas specific internal energy, $P$ the local gas pressure and ${\cal P} \, {\equiv} \, \sqrt{\gamma} P$ the pressure in the observer frame. 
The internal energy and gas pressure are linked by a polytropic equation of state, that is $\Sigma\varepsilon = \frac{P}{\gamma'-1}$ with $\gamma'=5/3$. 
Finally, the entropy is ${\cal S} \equiv \frac{3}{2} \ln(P/\Sigma^{\gamma'} )$.
In the equations above, the BBH spacetime is encrypted in the lapse function $\alpha$, the spatial metric tensor $\gamma_{ij}$ and the shift vector $\beta^i$.}

\subsection{Binary black hole spacetime}
\label{sec:metric}

{The BBH spacetime is described with the so-called {\lq}Near Zone{\rq} (NZ) metric \citep{ireland_inspiralling_2016}, that we implemented and tested (see \citealt{mignon-risse_impact_2023}; this same metric was also used in \citealt{noble_circumbinary_2012}).
The NZ metric is valid at distance $r_i \, {\ge} \, 10 M_i$ from the $i-$th BH and is defined as }
\be
g_{\mu\nu}^\mathrm{NZ} = \eta_{\mu\nu} + h_{\mu\nu}^\mathrm{NZ}
\ee
{where 
$\eta_{\mu\nu}$ is the (flat) Minkowski metric and $h_{\mu\nu}^\mathrm{NZ}$ is a 2.5 post-Newtonian (PN) metric perturbation.
This approximation relies on an expansion in the slow-motion regime ($v/ \mathrm{c}  \, {\ll} \, 1$, with $v$ the BH's velocity) and weak-field ($\mathrm{G} M/(r \mathrm{c}^2) \, {\ll} \, 1${, with $r$ the separation \textit{or} the distance to each BH}) limit.
From $g_{\mu\nu}^\mathrm{NZ}$, $\alpha$, $\gamma_{ij}$ and $\beta^i$ are obtained and their spatial derivatives are computed numerically.
By prescribing a GR metric, the gas self-gravity and gravitational feedback onto the BBH is neglected; thus, the disc mass is assumed to be negligible compared to the BBH mass.
Therefore, and as in any non-radiative GR(M)HD simulation, length and time units scale with the BBH mass {$M$} while the gas mass is scale-free.
The metric components depend on the BBH's mass ratio and spins, which are input parameters, and on their positions and velocities, which are provided separately by resolving the equation of motion.
}

\subsection{Equation of motion of the BBH}
\label{sec:eom}

\begin{figure}
\centering
\includegraphics[width=\columnwidth]{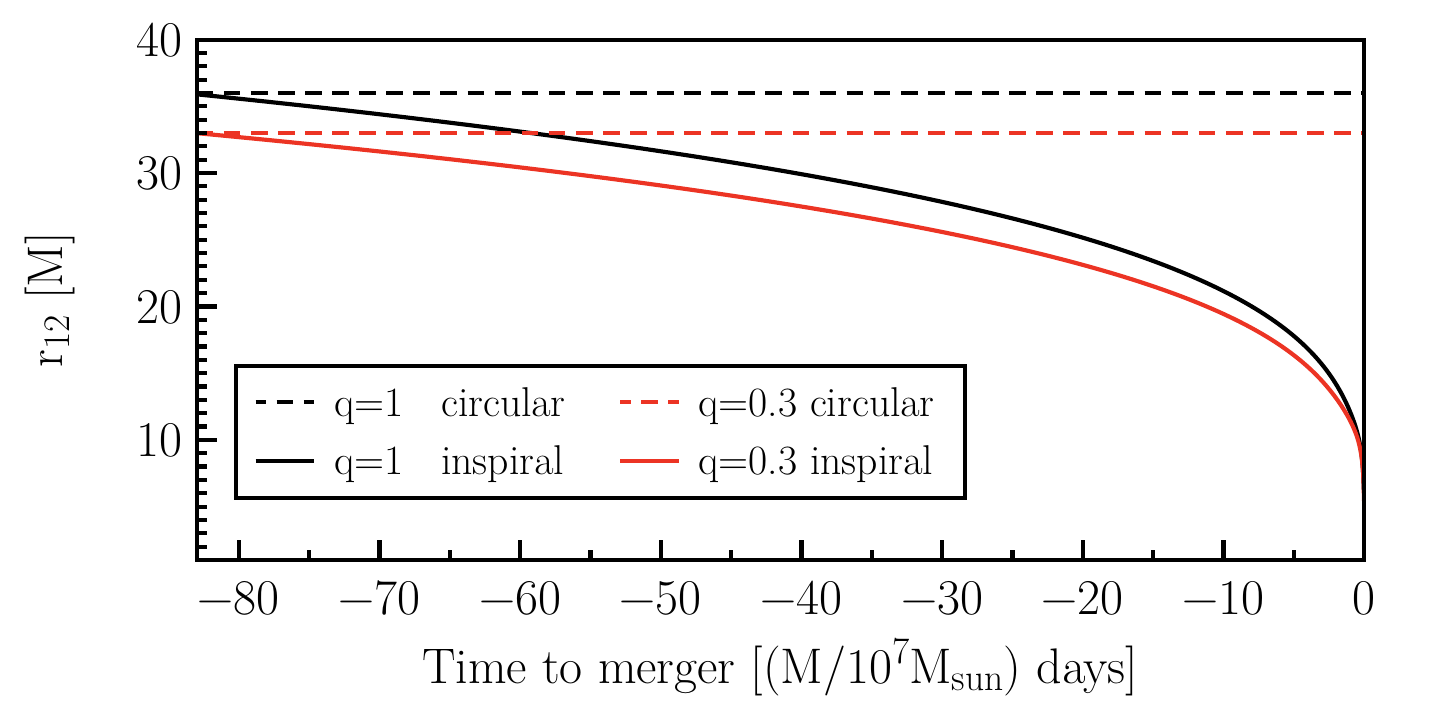}
\caption{Temporal evolution of the orbital separation ${r_{12}}(t)$, in units of $M$. The $x-$axis indicates the time to merger for the inspiral runs, chosen to be the same for $q\, {=} \, 1$ and $q\, {=} \, 0.3$.}
\label{fig:bfit}
\end{figure}

 {Since the BBH metric depends on their positions and velocities, we include the 3.5PN inspiral equation of motion for quasi-circular orbits (for more details, see \citealt{mignon-risse_impact_2023})}. 
 {This} requires one to obtain ${r_{12}}(t)$ from $t({r_{12}})$ (e.g. via Newton-Raphson inversion, as in \citealt{combi_superposed_2021}).
Because this is done at each timestep, and similarly, at each metric computation in the ray-tracing code, we turned to pre-computed fitting formulae of the PN separation and frequency (provided in App.~\ref{app:fit}).
Figure~\ref{fig:bfit} shows the evolution of ${r_{12}}(t)$ as a function of time in the different runs.
The initial separations are chosen to be $r_{12,0}\, {=} \, 36$~M for $q\, {=}\, 1$ and $r_{12,0}\, {=}\, 33$~M for $q \, {=} \, 0.3$ and the orbital periods are $\mathrm{P_{orb,0}} \, {=} \, 1411$~M and $1243$~M, respectively.
With this choice, the inspiralling time is the same in both runs.
The final separation is $r_{12,\mathrm{min}} \, {=} \, 8$~M.
{Let us specify that, unlike the study of \cite{franchini_emission_2024}, here the binary does not feel the gravitational effect of the gas.}

\subsection{Initial conditions in the circular-orbit runs}

{Inspiral runs are initialized with the data from circular-orbit runs.}
The{se circular-orbit} runs are initalized with an axisymmetric disc with the density profile
\be
\Sigma(r) \, {=} \, 0.5 \Sigma_0 \left( 1 - \tanh{\left( \frac{r-r_\mathrm{out}}{\Delta} \right) } \right) r^{-3/4} + \Sigma_\mathrm{min},
\ee
{where $\Sigma_0 \, {=} \,  10$ {(in arbitrary units)}, $P\, {=} \,  P_0 \Sigma^{\gamma'}$ with $P_0 \, {=} \, 1.8\times 10^{-3}$. 
The disc outer edge is set by the $\tanh$ function with $r_\mathrm{out}\, {=} \, 1000$~M, radial extent $\Delta \, {=} \, 500$~M and $\Sigma_\mathrm{min} \, {=} \, 10^{-4}$.
The initial azimuthal velocity is set so as to impose radial equilibrium in an equivalent Kerr metric (see e.g. \citealt{casse_impact_2017}) 
\be
v^\phi \, {=} \, \frac{ - \partial_r \beta^\phi \pm \sqrt{ (\partial_r \beta^\phi)^2 + 2 \alpha \gamma^{\phi \phi} \partial_r \gamma_{\phi \phi} (\partial_r \alpha + \alpha \frac{\partial_r P}{\Sigma h \Gamma^2}) } }{\alpha \gamma^{\phi \phi} \partial_r \gamma_{\phi \phi} },
\ee
as in \citealt{farris_binary_2011}.
Thus, the inner, low-density cavity is not present initially and will form due to the effect of the BBH gravity.
}

\subsection{Additional numerical aspects}
\label{sec:num}

To study the EM variability produced by the lump, one must resolve the CBD edge width with enough cells to capture lump formation and evolution\footnote{Insufficient resolution was found to result in the absence of lump.}.
Thus, we implemented a linear-log grid with constant {radial cell size} $\mathrm{dr}=0.067$~M for $r\, {<}\, 15$~M, after which $\mathrm{dr}$ increases logarithmically up to a maximal value $\mathrm{dr}\,{\approx}\,1$~M.
Our grid has $(n_r,n_\phi)=(1008,400)$ cells.

{Our simulations cover t}he inspiral timescale: $\tau_\mathrm{GW} \, {\varpropto} \,  r_{12,0}^4$ with $r_{12,0}$ the initial separation and $\tau_\mathrm{GW}(q)/\tau_\mathrm{GW}(q\, {=} \, 1)\, {\approx}\, (1+q)^2/4q\, \, {\gtrsim}\, 1.4$ for $q\, {=} \,0.3$.
Late decoupling (${r_{12}} \, {\sim} \, 15$~M) was studied in e.g. \citealt{noble_circumbinary_2012} but here we consider systems {in the inviscid (though with effective viscosity) limit, thus} decoupling earlier.
Because large-separation inspiral is prohibitive, we {have} set $r_{12,0}$ as already consistent with post-decoupling ({Sec.~\ref{sec:decoupling};} as in \citealt{farris_binary_2011}).
Inspiral runs are integrated in time from ${\approx}\,160 000\, \mathrm{M}\, {\sim}\,80 (M/10^7 \mathrm{M_\odot}) \, \mbox{days}$ to ${\sim} \, 100 (M/10^7 \, \mathrm{M_\odot}) \, \mbox{mn}$ before merger\footnote{This inspiral time is ${\sim}11$ and ${\sim}128$ times larger than in \cite{noble_circumbinary_2012} and \cite{farris_binary_2011}, respectively.}.

{We excised the inner cavity because we focus on the CBD while the NZ metric is not valid at a radius $r_i \, {\lesssim} \, 10M_i$ from the $i-$th BH.
The relevance of this choice, followed in a number of studies (\citealt{macfadyen_eccentric_2008}, \citealt{noble_mass-ratio_2021}) is supported by the finding that the great majority of the gas penetrating the inner region of radius ${\sim}{r_{12}}$ ends up in the BHs (or in their individual accretion structure) and does not escape this region \citep{tiede_how_2022}.}
Thus, to follow the inward motion of the fluid following the inspiral we implemented a moving inner boundary.
While all cells are initially part of the simulation, we have deactivated cells below the radius $r \, {<} \, r_\mathrm{NZ}=M_1 {r_{12}} + 10 M_2$, i.e. outside the NZ validity region.
The first four cells with $r \, {>} \, r_\mathrm{NZ}$ in the radial direction act as null-gradient boundaries on primitive variables, and as outflow boundaries for the velocity.
Since $r_\mathrm{NZ}$ decreases with time, we activate cells on-the-fly according to this same criterion. 

The final CPU cost of each of these simulations was about $300$~kCPUh.

\subsection{{Ray-tracing step}}
\label{sec:raytracing}

{In order to compute the EM observables, we perform the ray-tracing step with {\tt GYOTO} \citep{vincent_gyoto_2011}, in the same BBH metric as in the fluid simulation, also incorporating also the Far Zone metric, valid beyond a gravitational wavelength (again, see \citealt{mignon-risse_impact_2023} where tests are presented).
The gas is placed in the $z\, {=} \, 0$ plane and we assume that it is optically thick.
Null geodesics are integrated from the observer to the source until they reach the gas.
The gas density and velocity at the time of emission is interpolated between {\tt GR-AMRVAC} outputs.
We did not use the fast-light approximation, i.e. the metric is computed on-the-fly as photons propagate and the final image is composed of photons emitted at different times and therefore associated with different simulations' snapshots.}

{The emission is thermal and follows Planck's law.
We convert the code units to physical units by setting the {effective} temperature at $r_\mathrm{in} \, {=} \, 6$~M (the last stable orbit in Schwarzschild BH discs) to the observationally-motivated value $T_\mathrm{in} \, {=}\,0.1$~keV for a $10^5 \mathrm{M_\odot}$ \citep{lin_38_2013}.
In physical units, the temperature becomes $T(r,\phi) \, {=} \, T_\mathrm{in} \times (\Sigma(r,\phi) / \Sigma(r \, {=} \,r_\mathrm{in},\phi))^{\gamma'-1}$.
To scale this temperature to other masses we use the $\alpha$-disc model \citep{shakura_black_1973}, leading ultimately to $T_{in}\, {\varpropto} \, M^{-1/4}$ (e.g. \citealt{roedig_observational_2014}).}
The computing time associated with the purely ray-tracing step is comparable to that of the fluid simulation.

\section{Initial conditions for the inspiral: circular runs}
\label{sec:circ}

{The circular-orbit runs produce the initial data for inspiral.
They capture two} physical timescales of interest, despite their dependence on $r_{12,0}$ and $q$: the lump formation timescale, {followed by the} inspiral timescale.
The lump usually forms after tens of BBH periods $\mathrm{P_{orb,0}}$ (e.g. \citealt{mignon-risse_origin_2023}, \citealt{duffell_santa_2024}; {timescales longer than those} covered by pioneering works, \citealt{farris_binary_2011}, \citealt{bode_mergers_2012}), with $\mathrm{P_{orb,0}} \varpropto r_{12,0}^{3/2}$.
The restart time is thus {set to} $t\, {=} \,60 \, \mathrm{P_{orb,0}}$ for $q \, {=} \, 1$ {so the 
$m \, {=} \, 1$ mode, referred to as the lump, enters its dominating phase} (\citealp[see Fig.~4 of][]{mignon-risse_origin_2023}; and triggered via added random perturbations on $v^r$, of amplitude $0.002 \, v^\phi$) and $t\, {=} \, 16\, \mathrm{P_{orb,0}}$ for $q\, {=} \, 0.3$ because the lump appears earlier thanks to the symmetry breaking.
\\

Let us describe the flow morphology in these runs.
The initial setup does not contain a low-density cavity, which forms self-consistently because of the binary torques driving gas accretion or expulsion, at radius below ${\sim}2\, r_{12,0}$.
The morphology of the flow consists of two spiral streams, spiral arms in the CBD and the so-called overdense lump orbiting at the inner edge of the CBD, similar to previous studies (e.g. \citealt{dorazio_accretion_2013}).
{These structures are of main interest for this study as they provide angular momentum transport in the inviscid CBD.}

\subsection{Angular momentum transport and effective viscosity}
\label{sec:visco}

\begin{figure}
\centering
\includegraphics[width=\columnwidth]{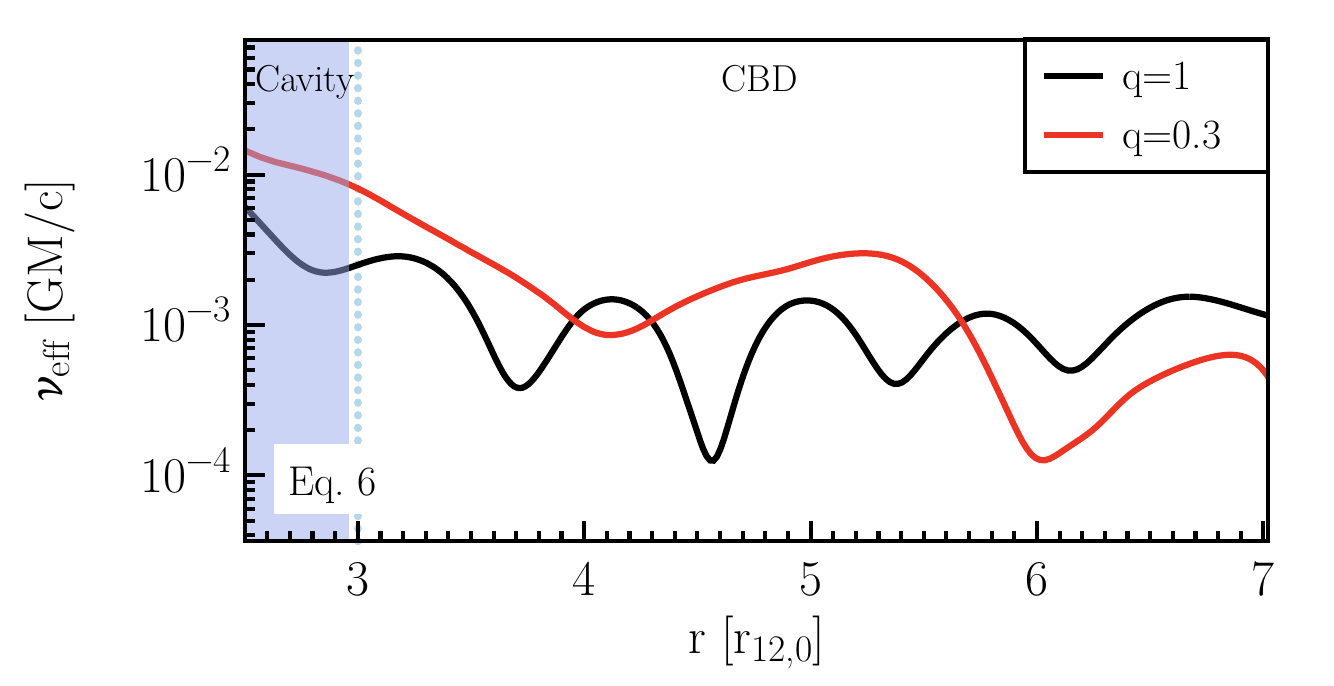}
\includegraphics[width=\columnwidth]{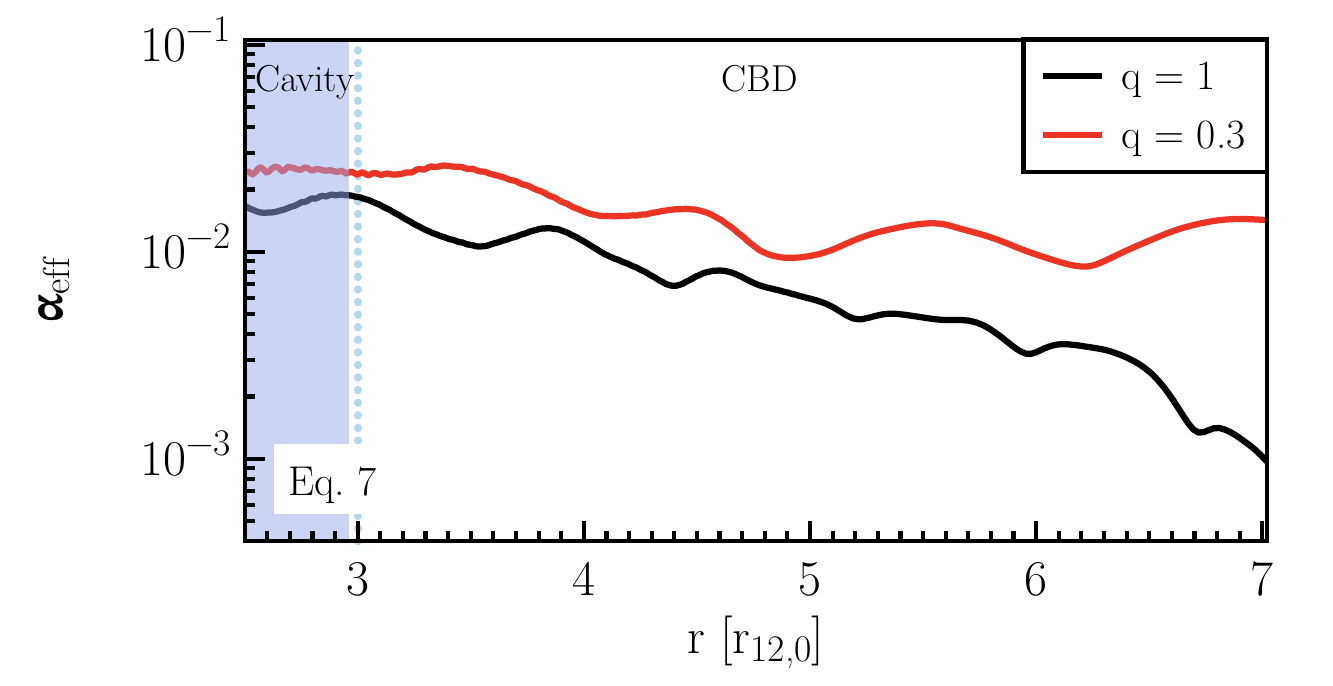}
\caption{ Radial profile of the time-averaged effective {kinematic (top panel, Eq.~\ref{eq:nueff}) and $\alpha$ (bottom panel, Eq.~\ref{eq:alphaeff})} viscosit{ies} (see the main text {for the two different methods of calculation}) at the time of starting the inspiral.
The vertical dotted line {is} located at $r \, {=} \, 3\, r_{12,0}$ to help locating the bulk of the CBD.}
\label{fig:nueff}
\end{figure}

{The notion of binary-disc decoupling is generally associated with the CBD $\alpha$-disc viscosity \citep{armitage_accretion_2002}.
In these inviscid simulations, the transport of angular momentum is, on top of the gravitational stress from the BBH, driven by hydrodynamical (Reynolds) stress due to the non-axisymmetries: spiral density waves, shocks and the lump (possibly associated with the Rossby wave instability; \citealt{mignon-risse_origin_2023}, \citealt{cimerman_gravitational_2024}). }

{In a classical -- Keplerian, axisymmetric, time-independent -- accretion disc, one can compute an effective kinematic viscosity
as $\nu_\mathrm{eff} \, {=} \,  -2/3 r v_r$ (e.g. Eq.~5.28, \citealt{frank_accretion_2002}).
To account for the flow's non-axisymmetry, we compute a density-weighted average as}
\be
\nu_\mathrm{eff} \, {=} \,  - \frac{2 \, r}{3} \, \frac{ {\langle} \, \Sigma v_r \, {\rangle_\phi} }{  {\langle} \, \Sigma \, {\rangle_\phi} }
\label{eq:nueff}
\ee
{ where ${\langle} . {\rangle}_\phi$ indicates the azimuthal average.
Then, we time-average $\nu_\mathrm{eff}$ over a timescale exceeding several binary orbits and more than one lump's orbit, around the time of starting the inspiral.}

The resulting radial profile of $\nu_\mathrm{eff}$ is shown in Fig.~\ref{fig:nueff}.
The vertical dotted line is located at $r\, {=} \, 3\, r_{12,0}$ to visually distinguish the CBD ($r\, {>} \, 3\, r_{12,0}$), beyond the localized, strongly non-axisymmetric impact from the lump; we will refer to this as the bulk of the CBD.
We find $\nu_\mathrm{eff} \,  {\sim} \, 10^{-4} \, {-} \, 5\times 10^{-3}  \mathrm{GM/c}$ in the bulk of the CBD, with an average of $\nu_\mathrm{eff} \, {\approx} \,  10^{-3} \mathrm{GM/c}$.
Meanwhile, the innermost regions have $\nu_\mathrm{eff}$ increasing up to $ {\sim} \, 0.1 \, {-} \, 1 \, \mathrm{GM/c}$ which we associate to the gravitational stress from the BBH (e.g. \citealt{tiede_gas-driven_2020}).
For comparison with other works, we can compute the equivalent (effective) $\alpha$ parameter. 
For a viscous, Keplerian disk, $\nu \, {=} \alpha (H/r)^2 \sqrt{GMr}$ \citep{frank_accretion_2002}{, with $H/r$ the disc aspect ratio} so we obtain $\alpha_\mathrm{eff}$ from $\nu_\mathrm{eff}$.
Fixing $H/r{\,{\approx}\,0.25}$ {as found at ${r\, {=} \, 3\, r_{12,0}}$}, this gives $\alpha_\mathrm{eff} \, {{\approx} \, 8 \times 10^{-3}}$ there.
As one could expect, this corresponds to the low-viscosity end among previous studies (see Tab.~\ref{table:codes}).

{The previous estimate of $\nu_\mathrm{eff}$, and the associated $\alpha_\mathrm{eff}$, remains based on standard (Keplerian, axisymmetric, time-independent) accretion disc theory. 
A more general approach consists in computing the ratio between the hydrodynamical stress and the thermal pressure.
We do so using the relativistic formalism of e.g. \cite{noble_circumbinary_2012} (see also \citealt{jiang_super-eddington_2019}), i.e. computing}
\be
\alpha_\mathrm{eff} = \frac{{\langle \langle}R^r_\phi {\rangle \rangle} + {\langle \langle}A^r_\phi {\rangle \rangle} }{ {\langle \langle}P {\rangle \rangle} },
\label{eq:alphaeff}
\ee
{ where $R^r_\phi$ and $A^r_\phi$ are Reynold's stress and advection flux of angular momentum (their sum makes up the hydrodynamical stress, denoted $T_\mathrm{H \, \phi}^{\, \, r}  $ in \citealt{noble_circumbinary_2012}), respectively, and ${\langle \langle}. {\rangle \rangle}$ is a time- and azimuthal-average operator.
In the bulk of the CBD, we get $\alpha_\mathrm{eff}$ ranging from $10^{-3}$ to a maximal value of $2 \times 10^{-2}$ near the CBD inner edge.
It is indeed smaller than usual values used in the literature (Tab.~\ref{table:codes}).}

Now th{e} effective viscosity is defined and quantified, we compute the expected separation at decoupling and show our setups are compatible with post-decoupling.

\subsection{Orbital separation at decoupling}
\label{sec:decoupling}

We follow the work of \cite{dittmann_decoupling_2023} to compute the separation at binary-CBD decoupling and show our simulations' setup corresponds to the post-decoupling epoch.
This time is often defined from the merger timescale equaling the viscous timescale at the CBD cavity radius $R_\mathrm{cav}=\xi {r_{12}}$, with $\xi$ a constant.
The separation reads
\be
{r_{12,\tau}} = \left( \frac{4 \lambda \xi^2}{3} \frac{A}{\nu_\mathrm{eff}} \right)^{1/2},
\label{eq:b_tau}
\ee
with $\lambda \, {\sim} \,  4/3 \, {-} \, 2 $ is a factor intervening in the viscous timescale estimation, {and} $A\, {\equiv} \, 64 \mathrm{G}^3 M_1 M_2 \mathrm{M} / (5 \mathrm{c}^5)$.
This definition is not unique, however, and the time-to-merger evolves much more rapidly ($\tau_\mathrm{GW} \, {\varpropto} \, {r_{12,0}}^4$) than the viscous timescale so a criterion based on the velocities may be more justified (\citealt{armitage_accretion_2002}):
\be
{r_{12,\mathrm{vel}}} = \left( \xi \frac{2}{3} \frac{A}{\nu_\mathrm{eff} }  \right)^{1/2}.
\ee
{This second criterion was found to be in better agreement with hydrodynamical simulations by  \cite{dittmann_decoupling_2023}.}

Using the more conservative form of these criteria ($\xi \, {=} \, 2$, $\lambda \, {=} \, 4/3$), we get $r_{12,\tau}  \, {=} \, 150$~M 
and $r_{12,\mathrm{vel}} \, {=} \, 65$~M  
for $q \, {=} \, 1$, and $r_{12,\tau}  \, {=} \, 127$~M 
and $r_{12,\mathrm{vel}} \, {=} \, 55$~M for $q \, {=} \, 0.3$\footnote{
{
These are based on values of $\nu_\mathrm{eff}$ taken from Eq.~\ref{eq:nueff}. If instead we compute $\nu_\mathrm{eff}$ from the maximal value of $\alpha_\mathrm{eff}$ (Eq.~\ref{eq:alphaeff}), we get $95$~M, $41$~M, $91$~M, and $40$~M, respectively, leaving the conclusion of post-decoupling unchanged.
}
}. 
These radii are larger than the initial separation we chose.
Hence, our setups correspond to the post-decoupling epoch{, so we can study the survival of the CBD features, and in particular that of the lump}.

\section{Post-decoupling phase}

The decoupling corresponds, theoretically, to the time beyond which the disk inward velocity becomes smaller than the BBH one.
Consequently, the distance from each BH to the CBD edge necessarily increases, questioning the survival of the pre-decoupling CBD flow features.

\subsection{Impact on the flow morphology}
\label{sub:flow}

\begin{figure}
\centering
\includegraphics[width=0.49\columnwidth]{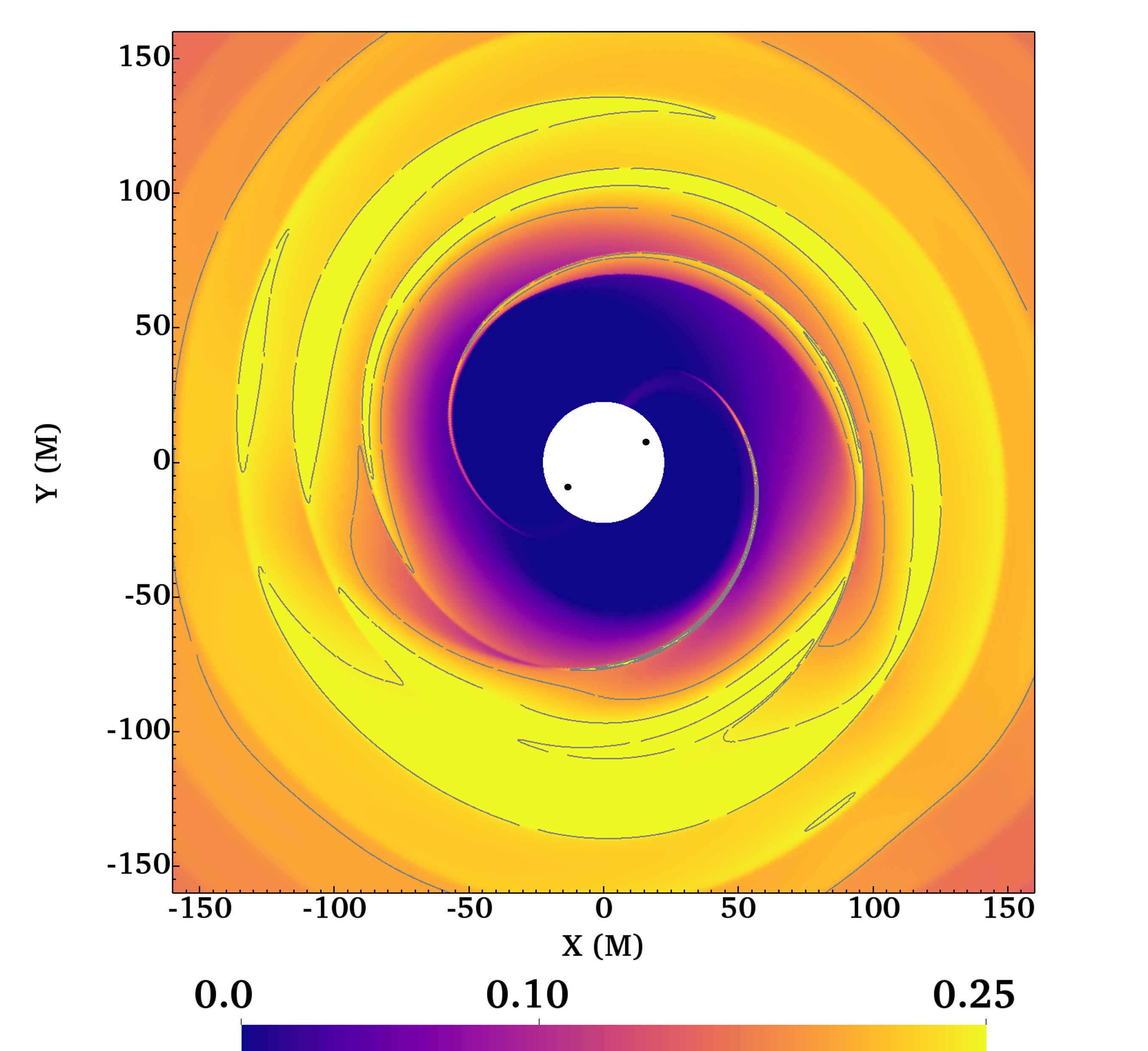} 
\includegraphics[width=0.49\columnwidth]{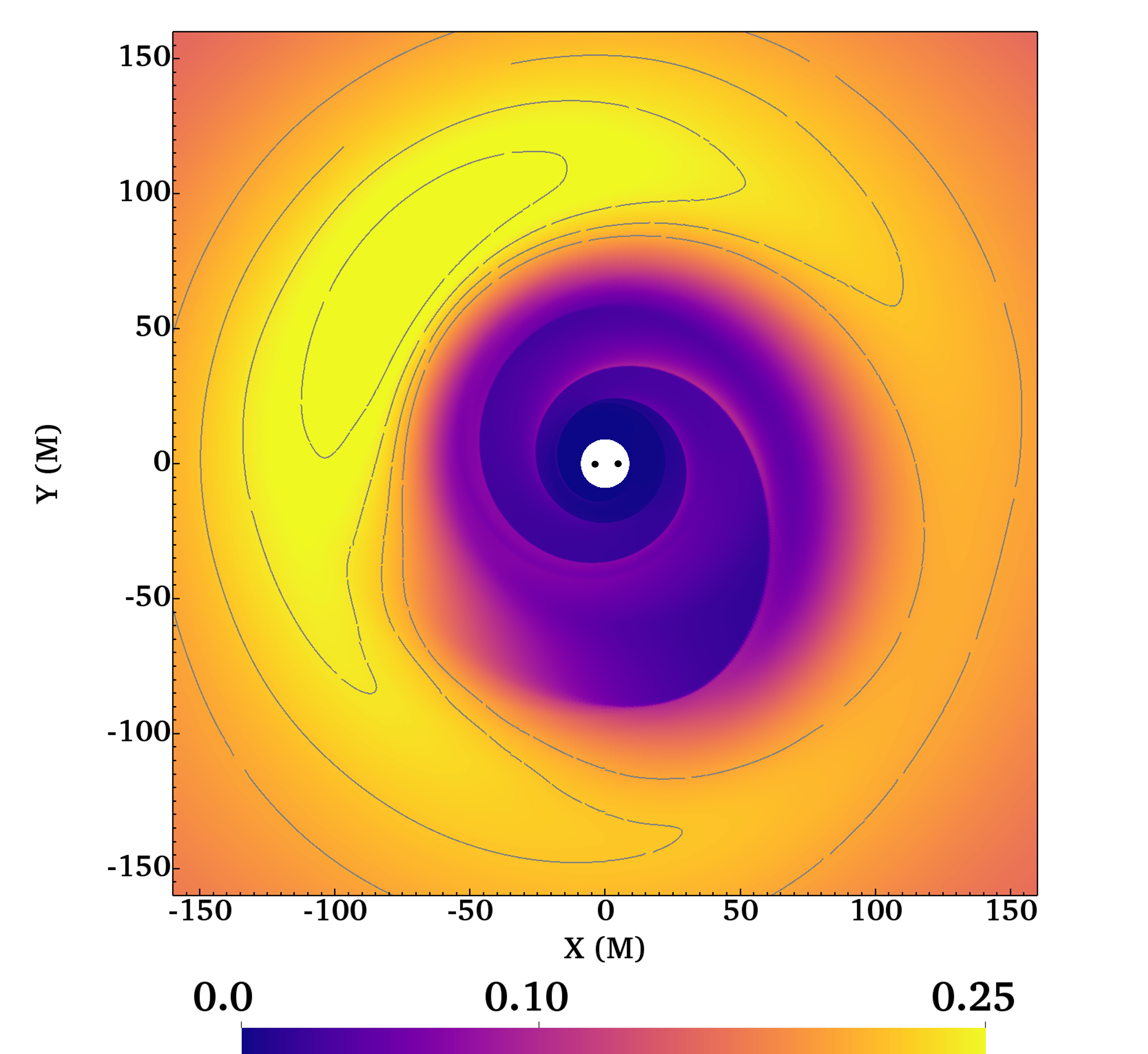} 
\includegraphics[width=0.49\columnwidth]{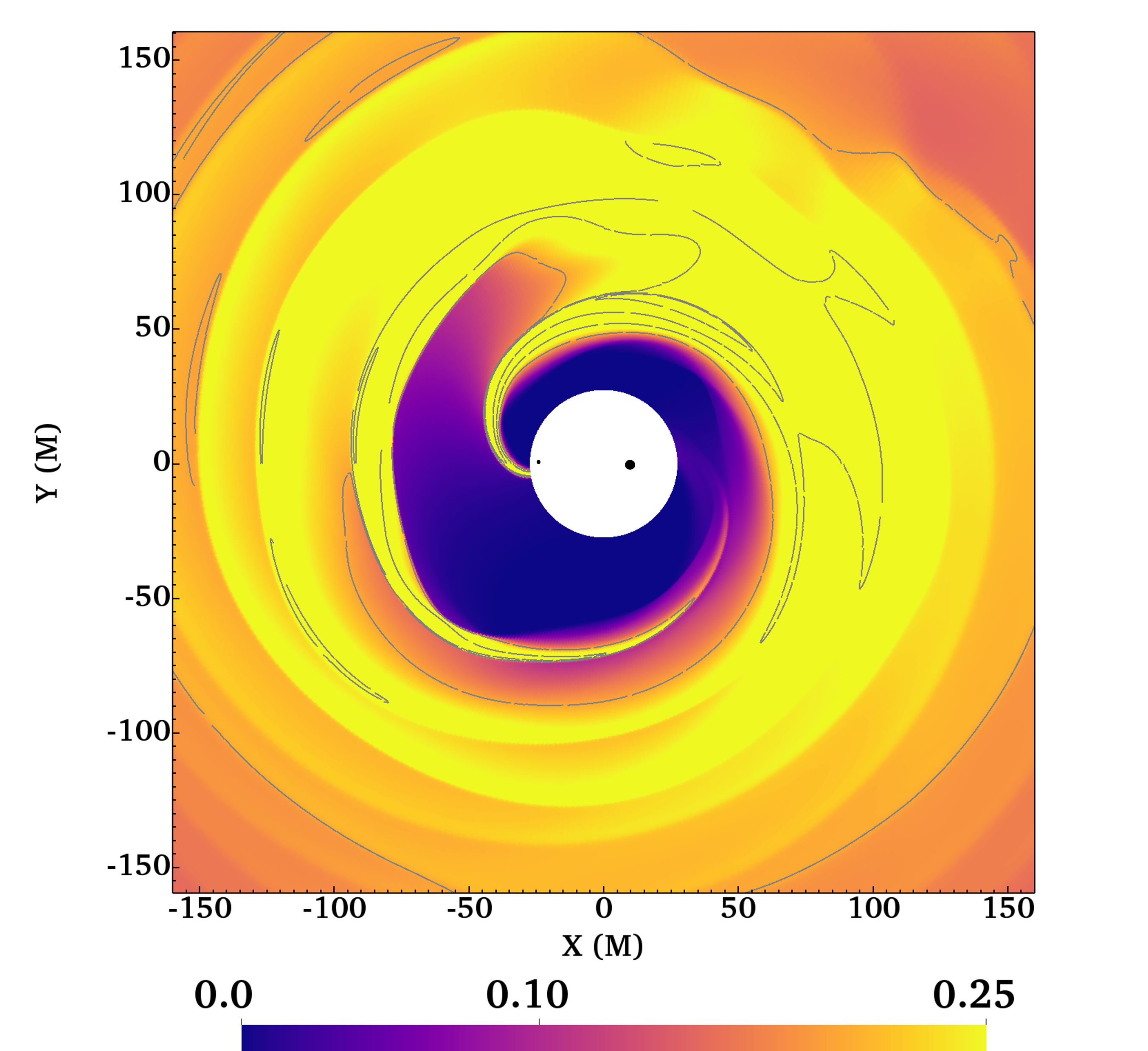} 
\includegraphics[width=0.49\columnwidth]{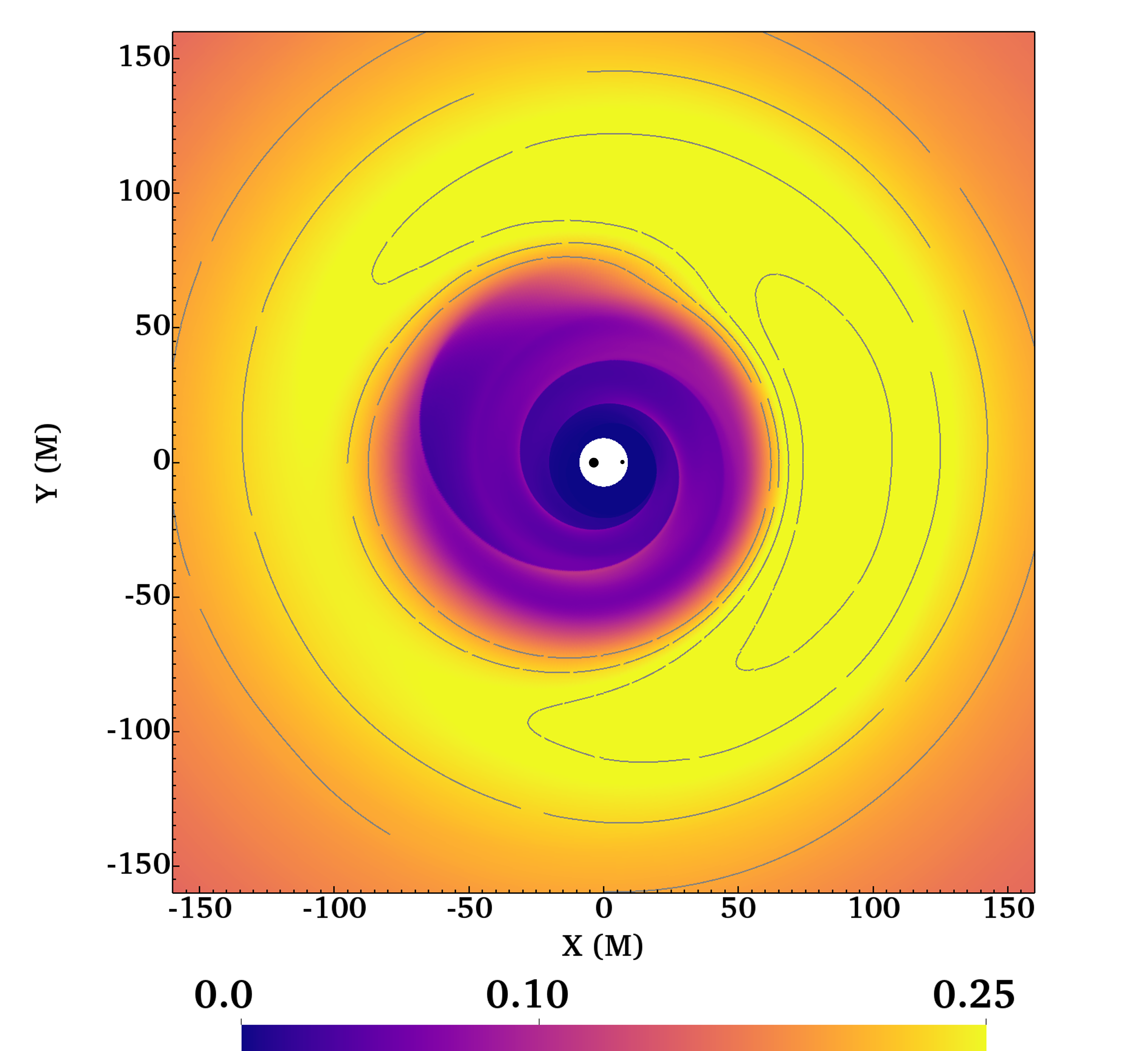} 
\caption{Density maps {in the inspiral runs} at ${\sim} 70 (M/10^7M_\odot)\, \mbox{days}$ (left column) and less than $1 (M/10^7M_\odot)\, \mbox{day}$ (near the end of the run, ${r_{12}} \, {=} \, 8\mathrm{M}$; right column) before merger for $q\, {=} \, 1$ (top) and $q\, {=} \, 0.3$ (bottom).
Density contours are overplotted in gray.
From the left to the right column, one can see how the grid inner boundary moves inward.
The BH {size} is increased by $10$ for visibility.}
\label{fig:rhomap_inspiral}
\end{figure}

Before studying the EM emission, we describe here the flow morphology.
{For this, the left panels of Fig.~\ref{fig:rhomap_inspiral} show the density maps ${\sim} 70 (M/10^7M_\odot) \, \mbox{days}$ before merger, corresponding to separations decreasing by ${\approx}1$~M with respect to the initial separation.}
The beginning of the inspiral runs exhibits all the features of the circular run{ (Sec.~\ref{sec:circ}), despite our inspiral simulations being in the post-decoupling regime from the beginning.}

{The right panels Fig.~\ref{fig:rhomap_inspiral} show the density maps less than $1 (M/10^7M_\odot) \, \mbox{day}$ before merger.} 
Importantly for observational consequences, the lump survive{s} to the decoupling stage and{, in our inviscid setup,} remains present at its pre-decoupling radius ${\sim}2\,{-}\,3 \, r_{12,0}$, not only for $q\, {=} \,1$ \citep{franchini_emission_2024} but also for $q\, {=} \, 0.3$.
{Apart from the lump's survival, the flow morphology is different (e.g. \citealt{dittmann_decoupling_2023}).
The initially low-density cavity is being partially filled with material, on top of which weak spiral arms are visible.
In the bulk of the CBD, the density distribution is much smoother, and only an $m\, {=} \, 1$ spiral wave, related to the lump, is visible instead of the $m\, {=} \, 2$ spiral waves mentioned previously.}
\\

\begin{figure}
\centering
\includegraphics[width=0.99\columnwidth]{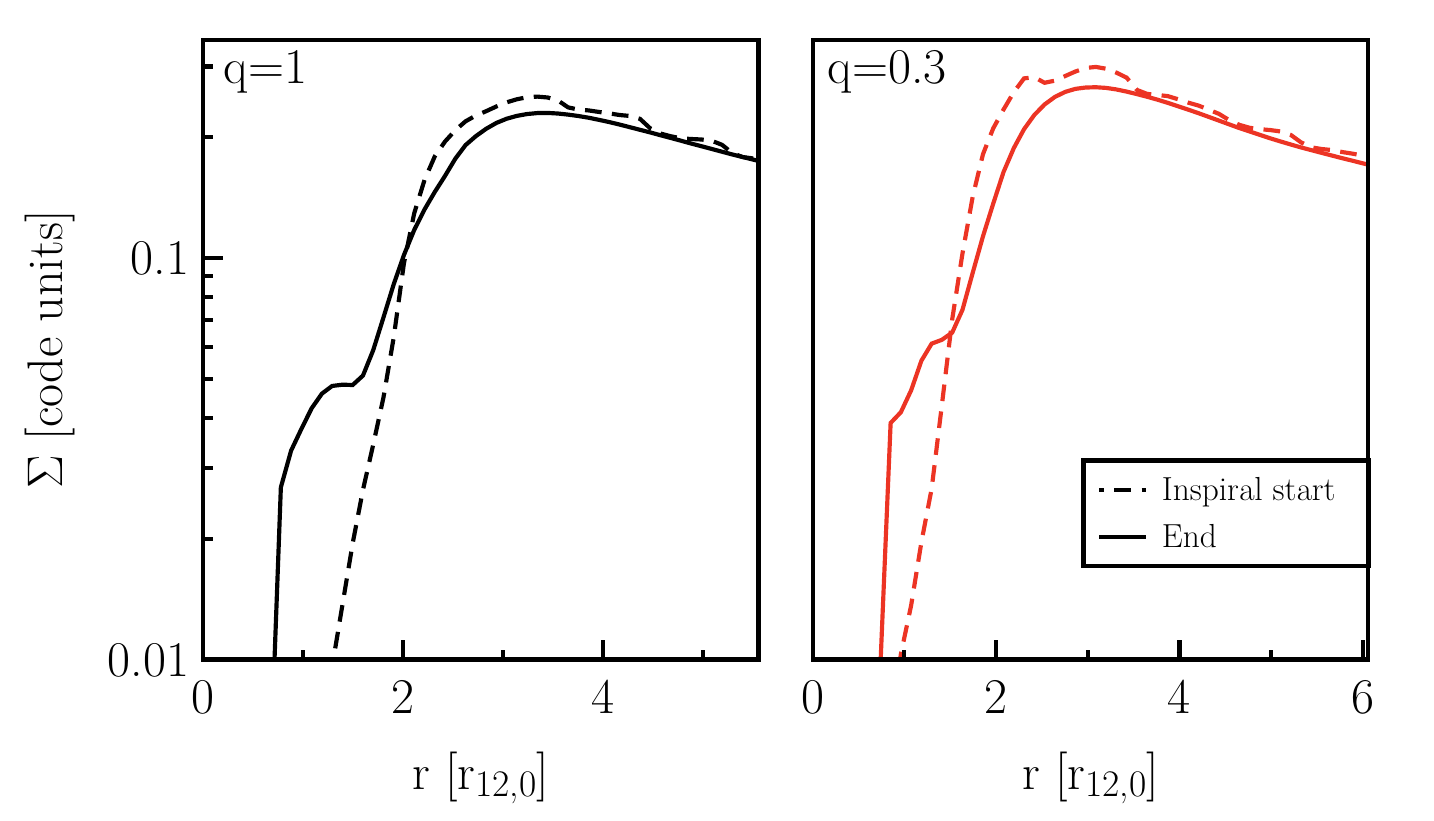}\\
\caption{{Azimuthally-averaged density profile at the start (dashed line) and end (full line) of the inspiral runs, for $q \, {=} \, 1$ (left) and $q \, {=} \, 0.3$ (right).}}
\label{fig:r_rho}
\end{figure}

\begin{figure}
\centering
\includegraphics[width=0.99\columnwidth]{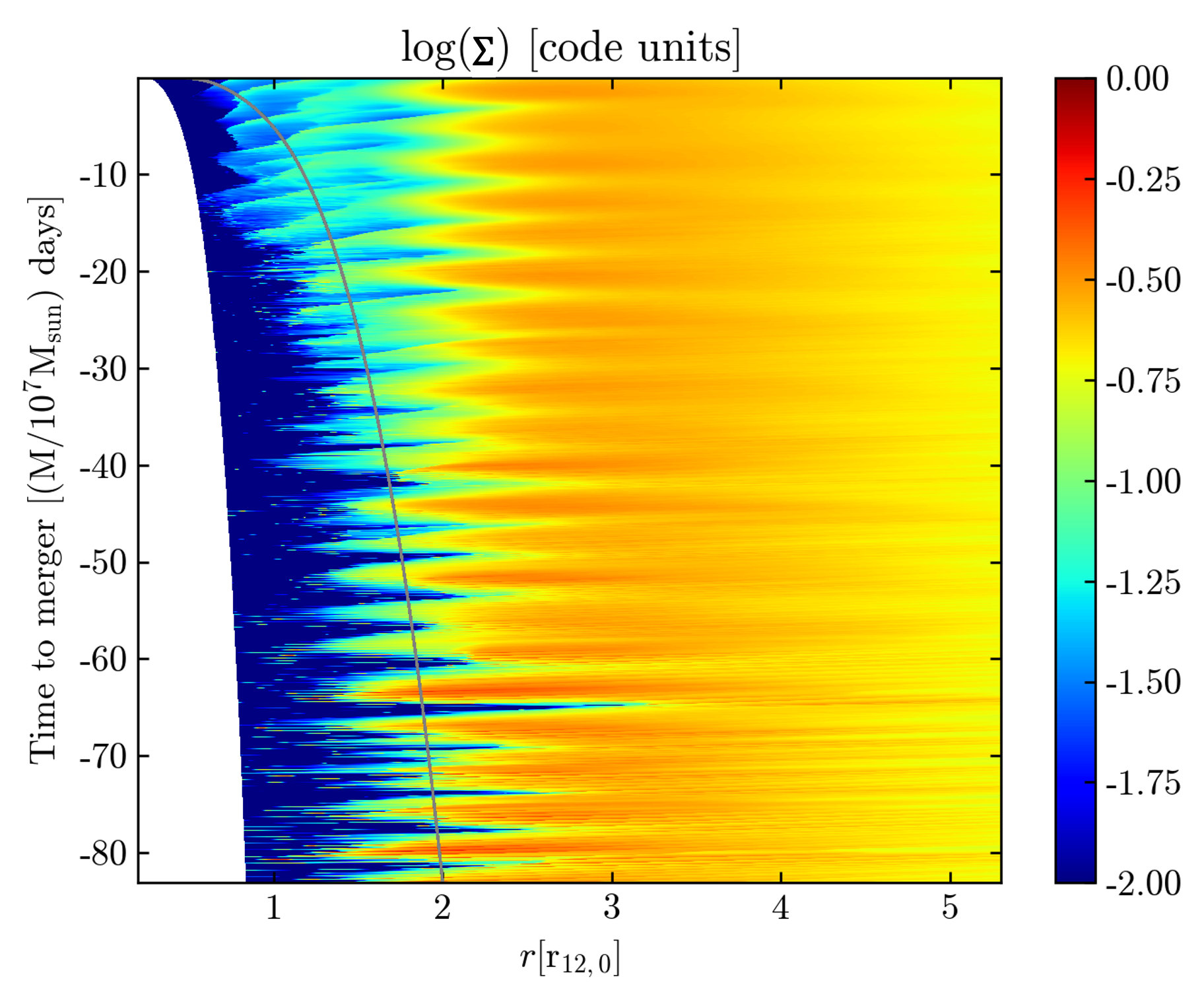}
\caption{Radius-time map of the density at $\phi \, {=} \, 0$ to show the evolution of non-axisymmetries post-decoupling{, }
for $q\, {=} \, 0.3$. {The gray curve indicates $2 \, r_{12}(t)$.}}
\label{fig:rtrho_inspiral}
\end{figure}

{Since the flow morphology has changed, we turn to following the evolution of the CBD inner edge.
Figure~\ref{fig:r_rho} shows the azimuthally-averaged density profile at the start (dashed lines) and at the end (full lines) of the inspiral runs.
In complement, Fig.~\ref{fig:rtrho_inspiral} shows the radius-time map of the density along the $\phi=0$ direction in order to show the time-evolution of gas non-axisymmetries.}
The {location of the maximal} CBD {density} 
stays at ${\sim}2\, r_{12,0}{\sim}70$~M, even in the late inspiral, while $2\, r_\mathrm{12,min} \, {=} \, 16$~M.
Meanwhile, 
 the CBD edge spreads inwards, filling partially the plunging region.
{T}his region contains ${\sim}4$ times more mass than in the control run at the same time.
{It is shown in Fig.~\ref{fig:r_rho} that this mass comes from the CBD densest region, suggesting it has been driven inward by the hydrodynamical angular momentum transport discussed in Sec.~\ref{sec:visco}.
This contrasts with the CBD recession reported in the inviscid study of \cite{farris_binary_2011}. 
We attribute this difference to our initial data, in which we evolved the circular-orbit runs over longer timescales to get these non-axisymmetric structures, responsible for angular momentum transport, to fully develop (Sec.~\ref{sec:circ}).
}
 {Some of the gas follows the BBH inspiral, indicated by the black curve in Fig.~\ref{fig:rtrho_inspiral}, but there is still a low-density cavity located around ${\sim}2\, r_{12}(t)$.}
 {Meanwhile, }the mass accretion rate through the inner boundary of the domain decreases progressively by about one order of magnitude {compared to its initial value and to the circular-orbit run at the same time}. 
 {This decrease is in qualitiative agreement with \cite{noble_circumbinary_2012} who also excised the cavity region, and with the lower (kinematic) viscosity run of \cite{franchini_emission_2024}{. It is a manifestation of the decoupling occurring.}
{{This is concomitant with} the {decreasing} gravitational torque from the BBH, {which is the source of non-axisymmetries ensuring the angular momentum transport in the disk}, as its distance to the CBD densest region has increased {\citep{farris_binary_2011}}.
We indeed showed in Fig.~\ref{fig:nueff} that it dominates the effective viscosity in the innermost regions.}
\\

Overall, binary-CBD decoupling impacts qualitatively the flow in the innermost regions but leaves the lump qualitatively unaffected{ for both $q\, {=} \,1$ and $q\, {=} \, 0.3$.
For $q\, {=}\, 1$, the lump's survival post-decoupling is in agreement with previous studies (\citealt{tang_late_2018}, \citealt{franchini_emission_2024}), and the evolution of the cavity and of the lump's position resemble the lower (kinematic) viscosity run of \cite{franchini_emission_2024}.}

\subsection{Electromagnetic emission}

Following our CBD peak temperature {prescription} (Sec.~\ref{sec:raytracing}, e.g. \citealt{roedig_observational_2014}), the SED peaks in the UV in the source frame for LISA-type sources ($M\, {=} \, 10^6 \mathrm{M}_\odot$) and in the optical for PTA-type sources and current targets of optical searches ($M{\sim}10^{8-10} \mathrm{M}_\odot$, e.g. \citealt{graham_systematic_2015}).
The {CBD} energy spectrum variation due to the inspiral is weak{ because the cavity filling is minor, compared to} the variability we now focus on.

\begin{figure}
\centering
\includegraphics[width=\columnwidth]{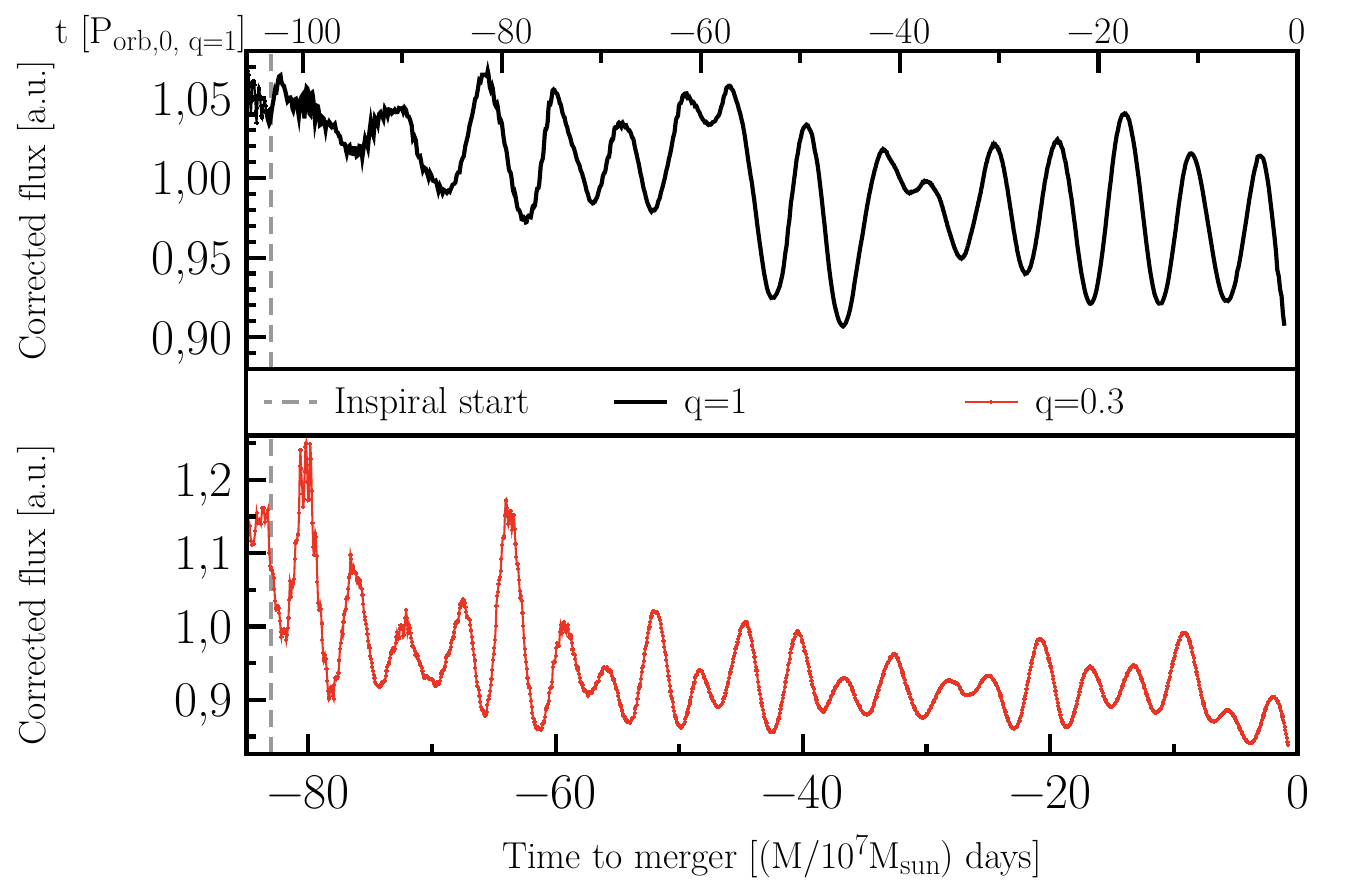}
\caption{Bolometric lightcurve for $q \, {=} \, 1$ (black) and $q \, {=} \, 0.3$ (red), with the flux normalized to its mean value.
The inspiral starting time is indicated by the grey line.
The time indicated on the $x-$axis is the time to merger, in days for $M\, {=} \, 10^7 \mathrm{M}_\odot$ (lower axis) and in units of $\mathrm{P_{orb,0}}(q\, {=} \, 1)$ (upper axis) for simplicity.}
\label{fig:LC}
\end{figure}

Since we are interested in the observational appearance of the post-decoupling CBD, we show in Fig.~\ref{fig:LC} the bolometric lightcurve for $q\, {=} \,1$ and $q\, {=} \,0.3$, for a disc inclination of $70 ^\circ$ (with $0 ^\circ$ being face-on).
The mean flux slightly decreases with time due to the limited mass reservoir (finite initial disc).
The {early} inspiral phase shows a {main} variability at $\mathrm{P_{lump}}\,  {\sim}\, 6\, \mathrm{P_{orb,0}}$.
The lump modulation is produced, mostly, by the relativistic beaming acting on the lump emission.
The most important result is the survival of the lump modulation until the end of the run, i.e. ${\sim}100 (M/10^7 \mathrm{M_\odot}) \, \mbox{minutes}$ before merger.
It remains qualitatively unchanged, with a similar period and a {slightly lower} amplitude.
{To further support this finding, we computed the Fourier transform of these LCs over the last $100 \, \mathrm{P_{orb,0}}$ divided into two epochs of equal duration to track any evolution of the LC variability.
The FFT power, multiplied by $\omega^2$ to enhance visibility over multiple frequency ranges, and normalized by its maximal value, is shown in Fig.~\ref{fig:fft}.
The peaks around ${\sim} \, 0.2\,  \Omega_\mathrm{orb,0}$ correspond to the lump's orbital frequency $\Omega_\mathrm{lump}$.
These peaks are present in both the early and late epochs, for both mass ratios.
Meanwhile, a smaller-amplitude peak at $\omega \, {\lesssim} \, 2\, \Omega_\mathrm{orb,0}$ is visible in the early epoch for $q\, {=} \, 1$ and is plausibly associated with the binary-lump beat, of expected initial frequency $2 \, \Omega_\mathrm{beat} \, {=} \, 2 (\Omega_\mathrm{orb,0} \, {-} \, \Omega_\mathrm{lump} ) \, {\sim} \, 1.7\,  \Omega_\mathrm{orb,0}$.
This method does not allow to capture the fast evolution of the beat frequency which follows the binary orbital frequency in the late inspiral.
}
Thus, we conclude on the survival of the lump modulation until the very last stages of the merger here{, in agreement with \citealt{tang_late_2018} for $q\, {=}\, 1$ and that we also showed for $q\, {=} \, 0.3$.}

\begin{figure}
\centering
\includegraphics[width=\columnwidth]{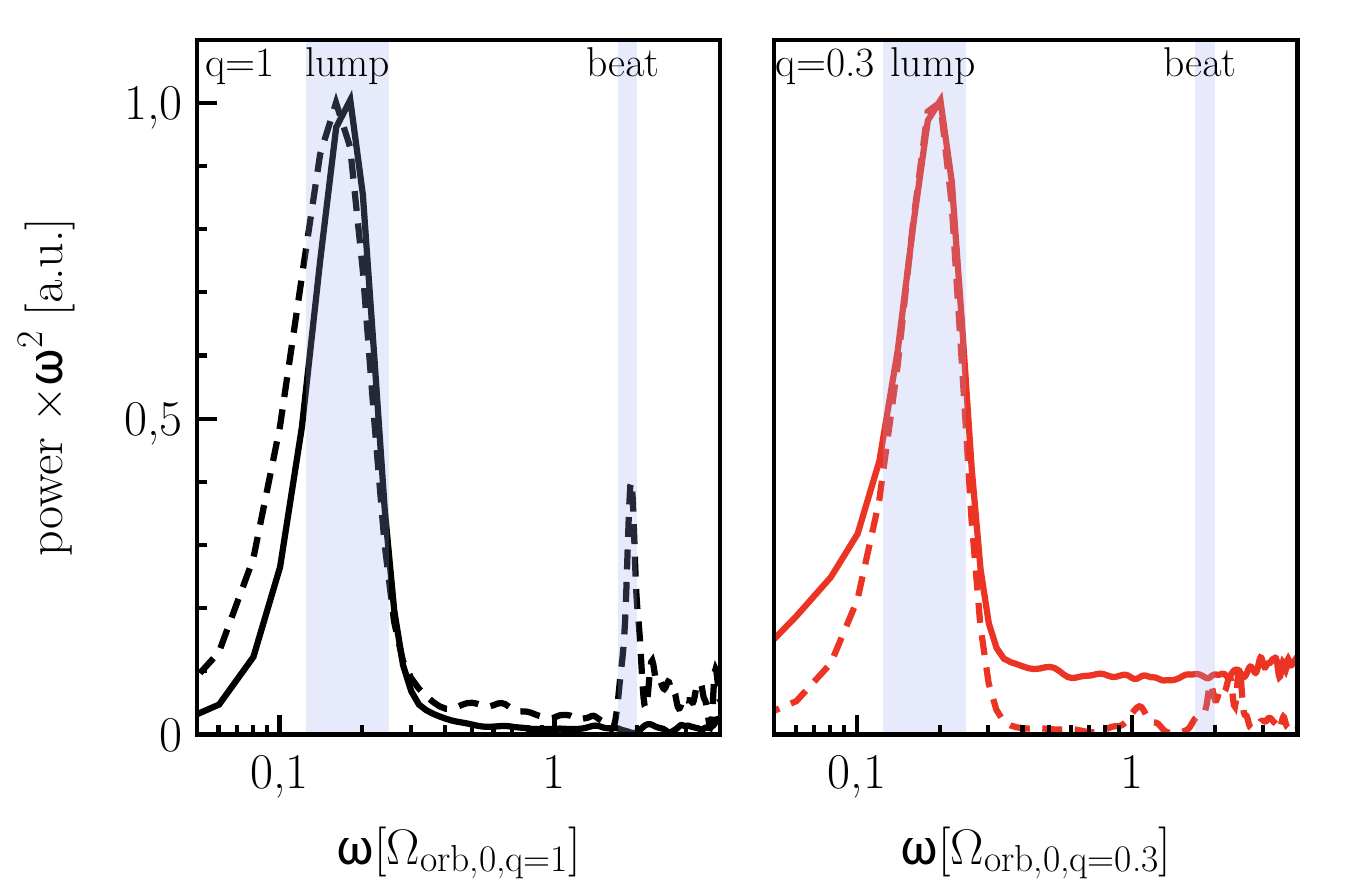}
\caption{
{Result of the Fourier transform of the LC for $q\, {=} \, 1$ (left panel) and $q\, {=} \, 0.3$ (right panel) over the last $50 \, \mathrm{P_{orb,0}}$ of the simulation (full line, referred to as the late epoch) and over the previous $50 \, \mathrm{P_{orb,0}}$ before this (dashed line, referred to as the early epoch), normalized to the peak value.}
}
\label{fig:fft}
\end{figure}

\section{Conclusion}

Most previous works on accreting, post-decoupling BBHs have focused on equal-mass binaries (e.g. \citealt{tang_late_2018}), on EM luminosity inherited from the accretion rate and/or mini-discs and therefore concluded to an EM dimming, mainly in the X-rays for $M \, {=} \, 10^6 \mathrm{M}_\odot$ BBHs typical of LISA sources (e.g. \citealt{krauth_disappearing_2023}).
{Angular momentum transport within the circumbinary disc was often included via an $\alpha$-viscosity or with the MRI.}
We took another approach here, focusing on {the post-decoupling evolution of inviscid circumbinary discs} and relativistic effects acting on {their} EM appearance, for $q \in \{ 0.3, 1 \}$ {non-eccentric binaries}. 

{First, we show that the hydrodynamical transport of angular momentum due to the circumbinary disc non-axisymmetries leads to an effective {alpha} viscosity {ranging from $10^{-3}$ in the bulk of the CBD to $2 \times 10^{-2}$ near the CBD inner edge.} 
As expected, this corresponds to the lower-end of the viscosity distribution among previous works.}
We {also find} that the accretion rate drops in the post-decoupling phase{, extending the conclusions reported in previous works (e.g. \citealt{krauth_disappearing_2023}, \citealt{dittmann_decoupling_2023}, \citealt{franchini_emission_2024}) to lower disc viscosity.}

{Our main result is the survival of the overdense lump post-decoupling, for both mass ratios, which modulates the EM flux via relativistic beaming.
It} survives until the end of our simulations when the approximate BBH metric is no longer valid, namely ${\sim}100 (M/10^7 \mathrm{M_\odot})\, \mbox{mn}$ before merger.
{In contrast with other studies incorporating additional angular momentum transport processes in the circumbinary disc, t}he lump radius and period are {indistinguishable from those} prior to decoupling.
{This result, together with previous ones, suggests the following: inviscid (here) and low-viscosity circumbinary disc (see the cold disc run of \citealt{franchini_emission_2024}) lead to the lump's survival at or close to its previous position (hence, orbital period), while more viscous circumbinary discs drive the lump's inspiral even after decoupling \citep{dittmann_decoupling_2023}, leading to its decreasing orbital period \citep{tang_late_2018}.
}

 {A} subset of the BBH candidates identified through their optical periodicity over $3-5$ cycles with $r_\mathrm{12,lump}\,{\lesssim}\,30$~M (e.g. SDSS J014350.13+141453.0;  \citealt{graham_systematic_2015}), a separation similar to the post-decoupling separation considered here, could already be post-decoupling {from an inviscid or low-viscosity circumbinary disc, unless they are still coupled to a more viscous disc}.
{The lump-related origin of their periodicity} 
can be tested with further optical follow-up.
The absence of GW signal in the PTA data associated with such a close-separation BBH, may rule-out the lump as the origin of the modulation, leaving accretion rate variability and Doppler boost over mini-disc emission{, thus at larger separation,} as the {main} binary-related origin of the modulation.

\section*{Acknowledgements}

RMR thanks A. Franchini and L. Mayer for useful discussions.
RMR acknowledges funding from Centre National d'Etudes Spatiales (CNES) through a postdoctoral fellowship (2021-2023). 
RMR has received funding from the European Research Concil (ERC) under the European Union Horizon 2020 research and innovation programme (grant agreement number No. 101002352, PI: M. Linares).
This work was supported by CNES, focused on Athena, the LabEx UnivEarthS, ANR-10- LABX-0023 and ANR-18-IDEX-000, and by the \lq Action Incitative: Ondes gravitationnelles et objets compacts{\rq} and the Conseil Scientifique de l'Observatoire de Paris. 
This work was also supported by the Programme National des Hautes Energies (PNHE) of CNRS/INSU, co-funded by CNRS/IN2P3, CNRS/INP, CEA and CNES.
The numerical simulations we have presented in this paper were produced on the platform DANTE (AstroParticule \& Cosmologie, Paris, France) 
and on the high-performance computing resources from GENCI - CINES (grant A0100412463) and IDRIS (grants A0130412463 and A0150412463).

\section*{Data Availability}
 
The data that support the findings of this study are available from the corresponding author, R.M.R, upon request.



\bibliographystyle{mnras}
\bibliography{Zotero} 



\appendix

\section{3.5 PN inspiral formulae}
\label{app:fit}

When rescaled to converge towards the 3.5PN value at large separations, the widely-used quadrupole formula \citep{peters_gravitational_1964} gives an error ${>}\,1\%$ (resp. ${>}\,5\%$) already at ${r_{12}} \,{\lesssim}\,24$~M (resp. ${r_{12}} \,{\lesssim}\,14$~M) for $q=1$, i.e. ${\lesssim}\,17 (M/10^7M_\odot) \mbox{days}$ (resp. ${\lesssim}\,50 (M/10^7M_\odot) \mbox{hours}$) before merger.   
It is, therefore, preferable to use higher-order (3.5 PN here) formulae for such post-decoupling simulations.
We parametrize ${r_{12}}$ and $\Omega$ as
\be
{r_{12}}(t') = a_1 t'^{1/4} + a_2 e^{-a_3 t'} ;
\Omega({r_{12}}) = \sqrt{\frac{1}{{r_{12}}(t')}} \left( 1+\frac{ \eta_\mathrm{eff} - 3}{2 {r_{12}}(t')} \right),
\ee
with $t'$ the time to merger, and $a_1$, $a_2$, $a_3$, and $\eta_\mathrm{eff}$ are all positive fitting parameters.
Physically, the first term in ${r_{12}}(t')$ is qualitatively similar to the quadrupole formula, on top of which we consider an exponential term that encapsulates the higher-order effects{ and that becomes increasingly important as the time to merger decreases.}
Then, $\Omega({r_{12}})$ is computed with the circular-orbit relation modified with the symmetric mass ratio $\eta$ replaced by an {\lq}effective{\rq} parameter $\eta_\mathrm{eff}$ that also absorbs the high-order effects.
Table~\ref{table:fit} gives the fitting parameters and the maximal fitting error for $q\in \{0.3, 1\}$\footnote{Upon request, we can provide a fit for different values of $q$ and even for spinning BBHs.}, valid down to ${r_{12}}\, {\approx}\, 8$~M. 
At ${r_{12}}\, {\approx}\, 8$~M the maximal error is reached but it is always below the percent level for ${r_{12}}\gtrsim 9~M$ (q=1), and ${r_{12}}\gtrsim 14~M$ (q=0.3), respectively. 
Comparable accuracy is reached for $\Omega$.

\begin{center}
\begin{table}
\caption{Fitting parameters for 3.5PN inspiral fit.}
\label{table:fit}  
\begin{tabular}{ | c | c | c | c | c | c} 
 \hline
 $q$ & $a_1$ & $a_2$ & $a_3$ & $\eta_\mathrm{eff}$ & Max. error [\%] \\  \hline
 $1$ & $1.838$ & $1.587$ & $0.00114$ & $0.602$ & $2$ \\  
 $0.3$ & $1.689$ & $1.974$ & $0.00076$ & $0.456$ & $3$ \\  
 \hline
\end{tabular}
\end{table}
\end{center}

\bsp	
\label{lastpage}
\end{document}